\documentstyle[epsfig,longtable]{aipproc}

\begin{document}
\title{Structure Functions at Very High $Q^2$ From HERA}

\author{Christopher M. Cormack$^*$}
\author{For the H1 and ZEUS Collaborations}
\address{$^*$Rutherford Appleton Laboratory, Chilton, Didcot, Oxford, OX11
0QX, United Kingdom}

\maketitle

\begin{abstract}
Measurements of Deep-Inelastic Neutral and Charged current
interactions are presented in lepton proton scattering at HERA.
The measurements are obtained from taken during 1996 to 1999 and
consists of $30 \, \rm{pb^{-1}}$ of $e^+p$ and $16\, \rm{pb^{-1}}$
of $e^-p$ data. The addition of the new high statistics electron
data with the positron data allows the first extraction of the
parity violating structure function $xF_3$ and tests of high-$Q^2$
electroweak effects of the heavy bosons $Z^0$ and $W$ are observed
and found to be consistent with the Standard Model expectation.
\end{abstract}

\section*{Introduction}

Deep--Inelastic scattering (DIS) of leptons off nucleons has
played a fundamental role in understanding the structure of
matter. HERA the first electron-proton ($ep$) collider ever built
is capable of investigating the two main contributions to DIS,
Neutral Current (NC) interactions, $ep \rightarrow eX$ and Charged
Current (CC) interactions, $ep \rightarrow \nu X$. In the Standard
Model a photon ($\gamma$) or a $Z^0$ boson is exchanged in a NC
interaction, and a $W^{\pm}$ boson is exchanged in a CC
interaction. DIS can be described by the four-momentum transfer
squared $Q^2$, Bjorken-$x$ and inelasticity $y$ defined as:
\begin{eqnarray}
Q^2 = -q^2\equiv(k-k')^2   \\
x=\frac{Q^2}{2p\cdot q}    \\
y=\frac{p\cdot q}{p\cdot k}
\end{eqnarray}
with $k(k')$ and $p$ being the four-momentum of the incident
(scattered) lepton and proton respectively. The centre-of-mass
energy $\sqrt{s}$ of the $ep$ interaction is given by $ s \equiv
(p+k)^2=Q^2/xy$.

In this paper measurements are presented of the NC and CC
cross-sections at high $Q^2 > 200 \, \rm{GeV^2}$ for both electron
and positron scattering from both the H1 and ZEUS experiments. The
integrated luminosity in the $e^+p$ data sample is $35.6 \,
\rm{pb^{-1}} \, \, (30 \, \rm{pb^{-1}})$ and in the $e^-p$ data
sample is $16 \, \rm{pb^{-1}} \, \, (16 \, \rm{pb^{-1}})$ for the
H1 (ZEUS) collaborations.

\section{Experimental Setup}
The H1 and ZEUS detectors are described elsewhere
\cite{h1},\cite{zeus}. Both detectors are nearly hermetic
multi-purpose apparatus built to investigate $ep$ interactions.
The measurements in both experiments rely primarily on the
calorimetry, tracking and the luminosity detectors.

In both experiments the Charged Current event selection is based
on the observation of large $P_T^{miss}$, which is assumed to be
the transverse momentum carried by the outgoing neutrino. For NC
events it is based on the identification of a scattered electron
(positron), further fiducial (NC) and kinematic cuts (CC and NC)
are then applied. Details of the selection procedure can be found
in \cite{zeus,h1nccc}.

\section{Cross Sections}
The electroweak Born-level NC DIS cross section
$d^2\sigma^{NC}/dxdQ^2$ for the reaction $e^{\pm} p \rightarrow
e^{\pm} X$ can be written as:

\begin{equation}
\frac{d^2\sigma_{Born}^{e^{\pm p}}}{dxdQ^2}=\frac{2\pi
\alpha^2}{xQ^4}\left[Y_+F_2(x,Q^2)\mp
Y_-xF_3(x,Q^2)-y^2F_L(x,Q^2)\right]
\end{equation}

For unpolarised beams, the structure functions $F_2$ and $xF_3$
can be decomposed taking into account $Z^0$ exchange as:

\begin{center}
\begin{eqnarray}
F_2 &\equiv & F_2^{em}- \upsilon \frac{\kappa_w
Q^2}{(Q^2+M_Z^2)}F_2^{\gamma
Z}+(\upsilon^2+a^2)\left(\frac{\kappa_w Q^2}{Q^2 +
M_Z^2}\right)F_2^Z \\

xF_3 &\equiv&  -a \frac{\kappa_w Q^2}{(Q^2+M_Z^2)}xF_3^{\gamma Z}
+ (2\upsilon a)\left(\frac{\kappa_w Q^2}{Q^2+M_Z^2}\right)^2
xF_3^Z
\end{eqnarray}
\end{center}
where $M_Z$ is the mass of the $Z^0$, $\kappa_w=1/(4\sin^2\theta_w
\cos^2\theta_w)$ is a function of the Weinberg angle $(\theta_w)$
and $\upsilon$ and $a$ are the vector and axial couplings of the
electron to the $Z^0$. For unpolarised beams, $F_2$ is the same
for electron and positron scattering, while $xF_3$ changes sign.

The NC ``reduced cross-section'' $\tilde{\sigma}$ is defined from
the measured as:
\begin{equation}
\tilde{\sigma}_{NC}(x,Q^2)\equiv
\frac{1}{Y_+}\frac{xQ^4}{2\pi\alpha^2} \frac{d^2\sigma^{NC}}{dxdQ^2}
\end{equation}

The Born double differential CC cross-section for $e^+p\rightarrow
\overline{\nu} X$ can be written in leading order QCD as:
\begin{equation}
\left(\frac{d^2\sigma^{e^+p}_{CC}}{dxdQ^2}\right)_{Born}=\frac{G_F^2}{2\pi
}\left( \frac{M_W^2}{M_W^2+Q^2}\right)^2
[(\bar{u}+\bar{c})+(1-y)^2(d+s)]
\label{eqn:cc}
\end{equation}
for $e^-p$ scattering the coupling is predominantly to the $u$ 
type quarks for $e^+p$ scattering the electroweak coupling is
predominantly to the $d$ type quarks.

\begin{equation}
\left(\frac{d^2\sigma_{CC}^{e^-p}}{dxdQ^2}\right)_{Born}=\frac{G_F^2}{2\pi}\left(\frac{M_W^2}{M_W^2+Q^2}\right)^2[u+c+(1-y)^2(\bar{d}+\bar{s})]
\end{equation}

The double differential CC ``reduced'' cross-section $\tilde{\sigma}_{CC}$ 
is defined as:
\begin{equation}
\tilde{\sigma}_{CC}=\frac{2\pi x}{G_F^2}\left(\frac{M_W^2+Q^2}{M_W^2}\right)^2 \frac{d^2\sigma_{CC}}{dxdQ^2}
\end{equation}

\section{Results and Interpretation}
\begin{figure}[Hpb]
\vspace*{-1.cm}
  \begin{center}
\flushleft{
    \epsfig{file=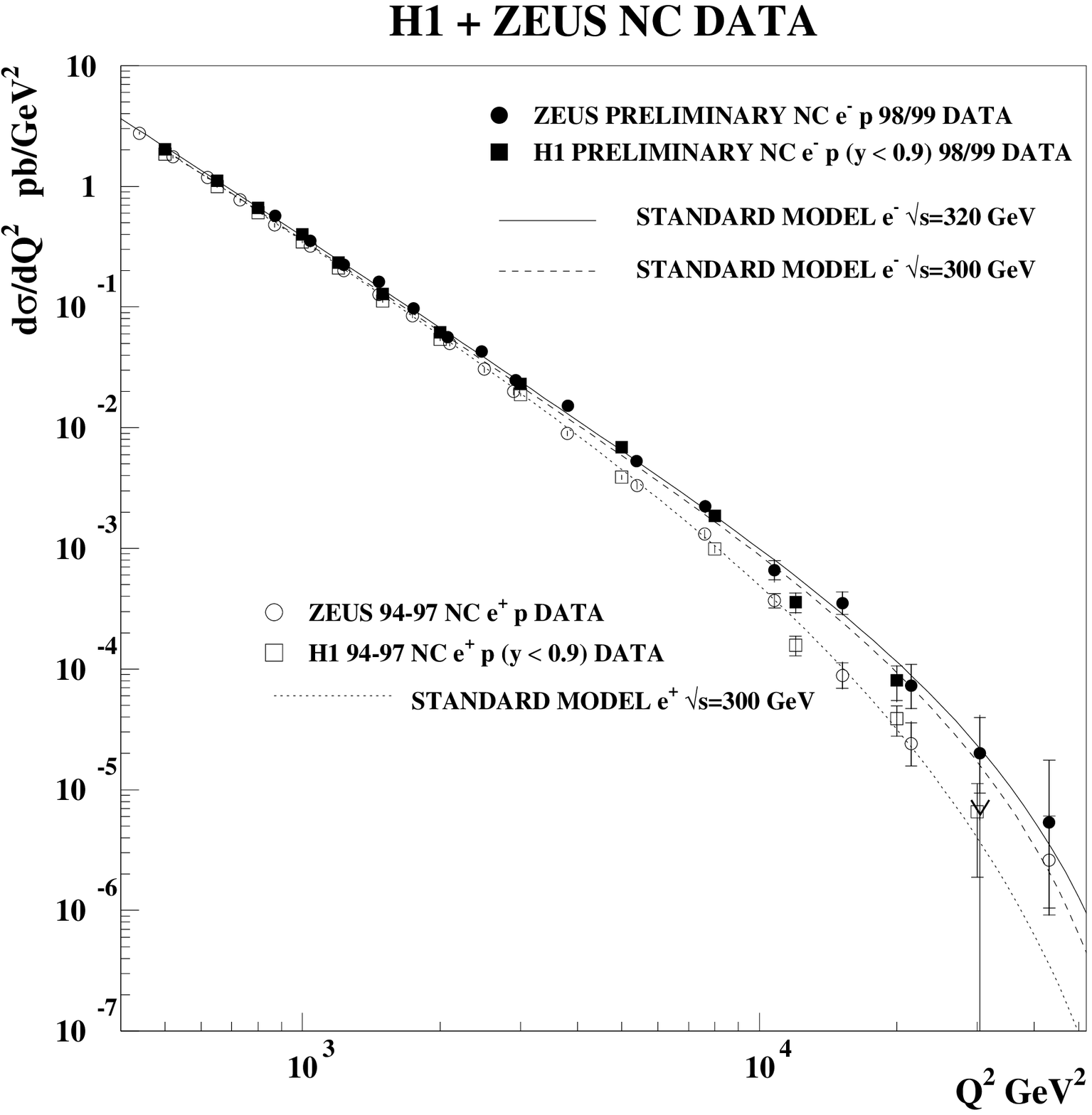,height=6.0cm, width=6.cm}}

    \epsfig{file=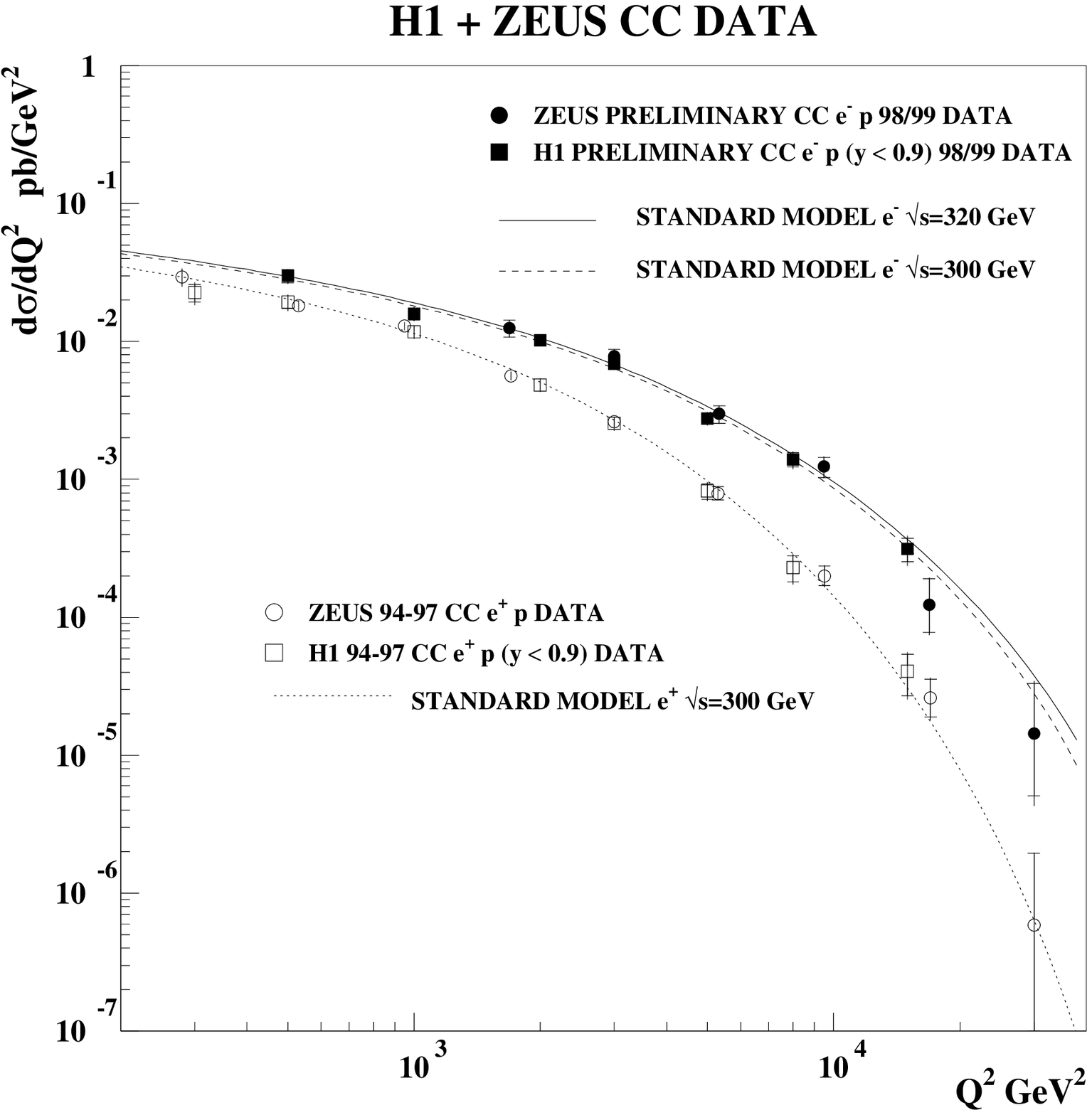,height=6.0cm, width=6.0cm}
\epsfig{file=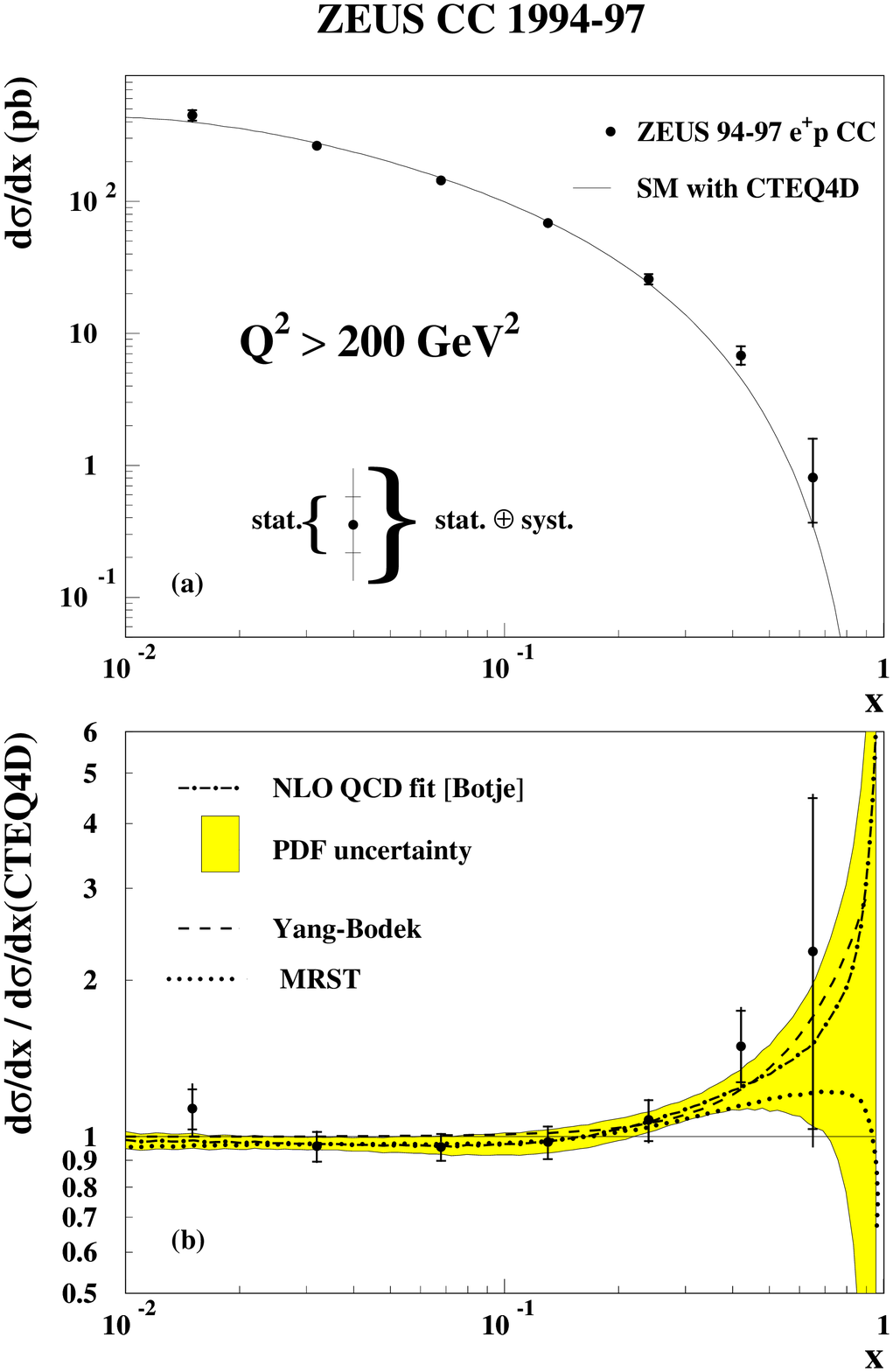,height=6.0cm, width=6.0cm, clip =true, bbllx=3,bblly=5,bburx=565, bbury=400}
\vspace*{-0.1cm}
    \caption{\footnotesize
      a) The high-$Q^2$ $e^{\pm}$ NC cross section $d\sigma_{e^{\pm}p}/dQ^2$
from the ZEUS and H1 collaborations along with the Standard Model predictions
using the CTEQ4D parton momentum distributions.
      b)The high-$Q^2$  CC cross section $d\sigma_{e^{\pm}p}/dQ^2$
from the ZEUS and H1 collaborations along with the Standard Model predictions
using the CTEQ4D parton momentum distributions.
      c) The ZEUS CC high-$Q^2$ structure function
$d\sigma/dx$ with the Standard Model prediction using CTEQ4D.
      d) The ratio of the measured ZEUS CC high-$Q^2$ structure function
$d\sigma/dx$ with the Standard Model prediction using CTEQ4D. Also shown
are the predictions from Yang-Bodek and the ZEUS NLO QCD fit.
    \label{fig:singdif}}
\end{center}
\end{figure}

The NC and CC single differential cross-sections and
$d\sigma_{NC}/dQ^2$, $d\sigma_{CC}/dQ^2$ are shown in fig
\ref{fig:singdif}, also shown are the Standard Model expectations
given from NLO QCD fits to the data by both the CTEQ \cite{cteq}
and the H1 \cite{h1qcd} collaborations. The Standard Model is seen
to give a good description of the data.

The measurement of he NC cross-section spans more than two orders
of magnitude in $Q^2$ and the cross-section falls with $Q^2$ by
about 6 orders of magnitude for both the $e^+p$ and the $e^-p$
cross section. For $Q^2$ values above $1000 \, \rm{GeV^2}$ the
$e^-p$ cross section can be seen to be significantly above the
$e^+p$ cross section. This is consistent with the Standard Model
picture in which one expects constructive interference of the
photon and $Z^0$ for $e^-p$ interactions and destructive
interference for $e^+p$ interactions.

Due to the propagator mass term and the different coupling the CC
cross-section is smaller than the NC cross-section and it falls
less steeply, by about 3 orders of magnitude, between $Q^2=300$
and $15\,000 \, \rm{GeV^2}$ for both the $e^-p$ and $e^+p$ cross
section. The shape and magnitude of the NC and CC cross-sections
are well described by the Standard Model expectation. The $e^-p$
cross section is at relatively low $Q^2$ to be approximately twice
as big as the $e^+p$ cross section, which is consistent with the
Standard Model picture in which the $u$ quark density probed is
expected to be approximately twice the $d$ quark density.

 Shown in fig. \ref{fig:singdif} is the single differential CC
 cross section $d\sigma_{CC}/dx$ for $Q^2 > 200 \, \rm{GeV^2}$ for
 $e^+p$ scattering from the ZEUS collaboration \cite{zeuscc}. The
 ratio of the measured cross section $d\sigma_{CC}/dx$. At high
 $x$ the $e^+p$ CC cross-section depends mainly on the $d$ quark density
 which is less constrained than the $u$ quark density. All data
 agree well with the Standard Model expectation using PDFs extracted from
CTEQ4D and those from the ZEUS NLO fit \cite{botje}. The
possibility of a larger $d/u$ ratio than previously assumed has
been of interest in recent years, for example see
\cite{yangbodek,melnitchouk}. Modification \cite{yangbodek} of
PDFs with an additional term $\delta(d/u)$ yields $d\sigma/dx$
close to the NLO QCD fit.

\begin{figure}[Hpb]
\vspace*{-1.cm}
  \begin{center}
    \epsfig{file=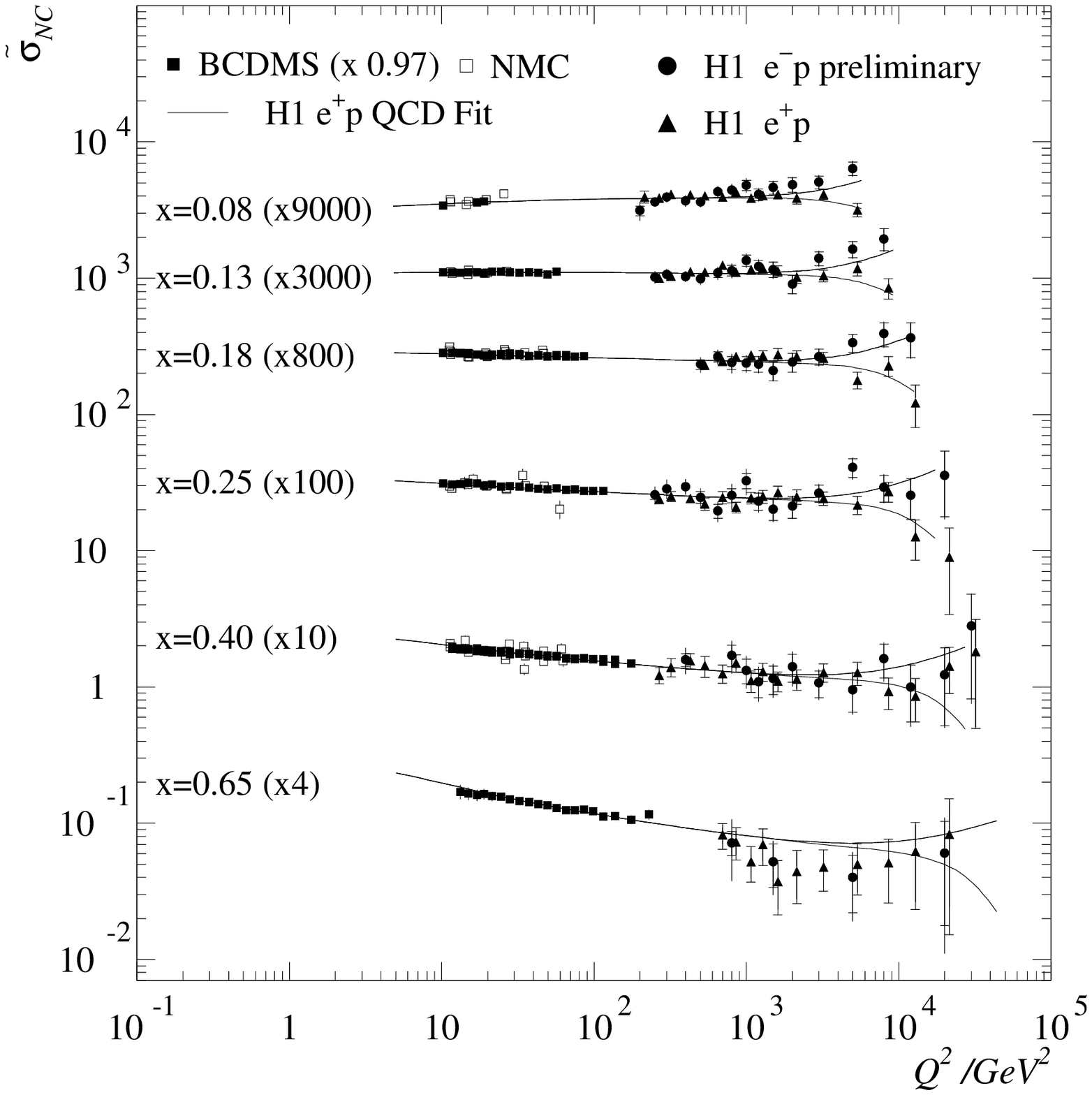,height=6.cm, width=6.cm}
\put(-140,175){\makebox(0,0){\footnotesize a)}}
    \epsfig{file=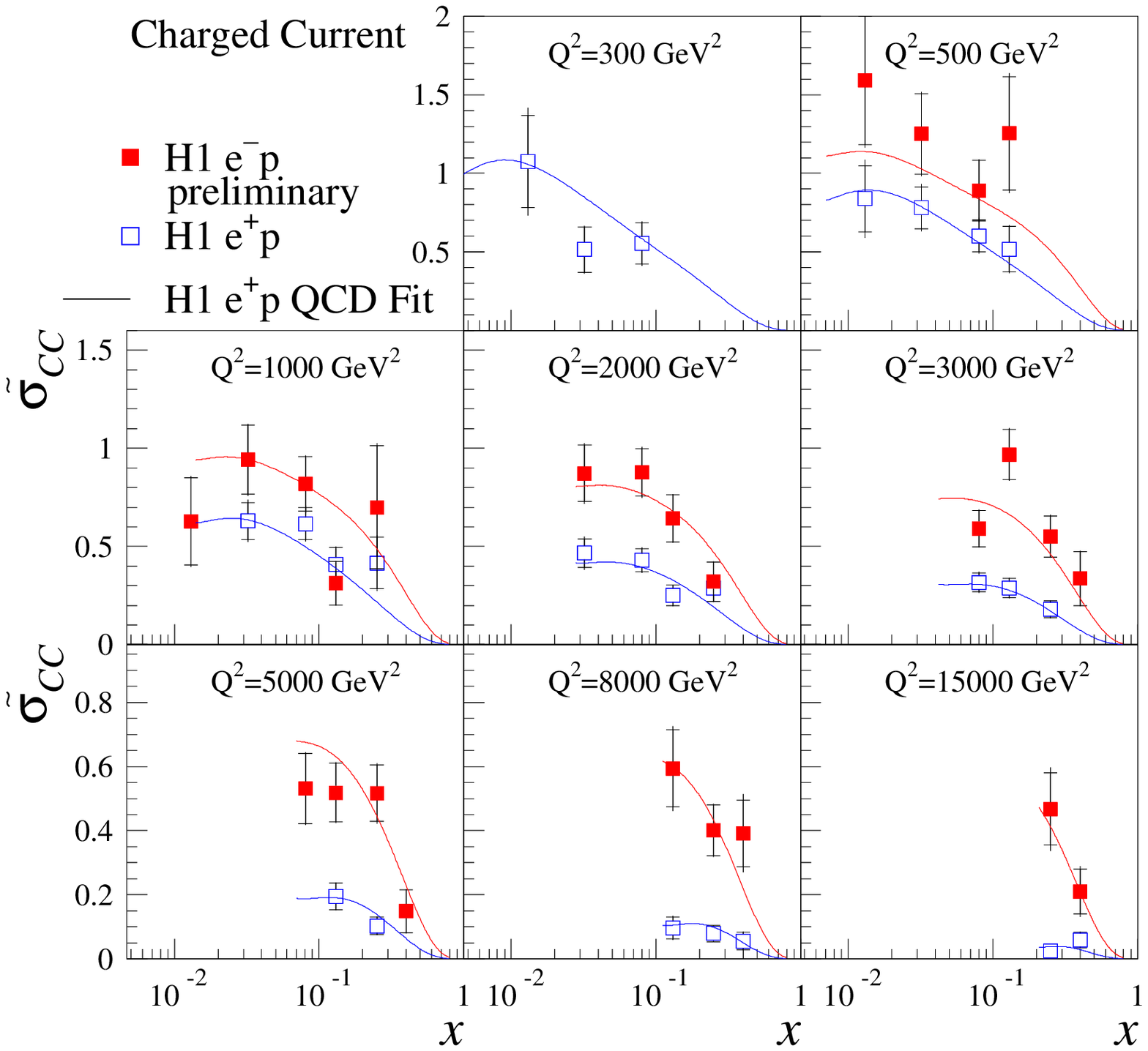,height=6.cm, width=6.0cm}
\put(-140,175){\makebox(0,0){\footnotesize b)}}

\epsfig{file=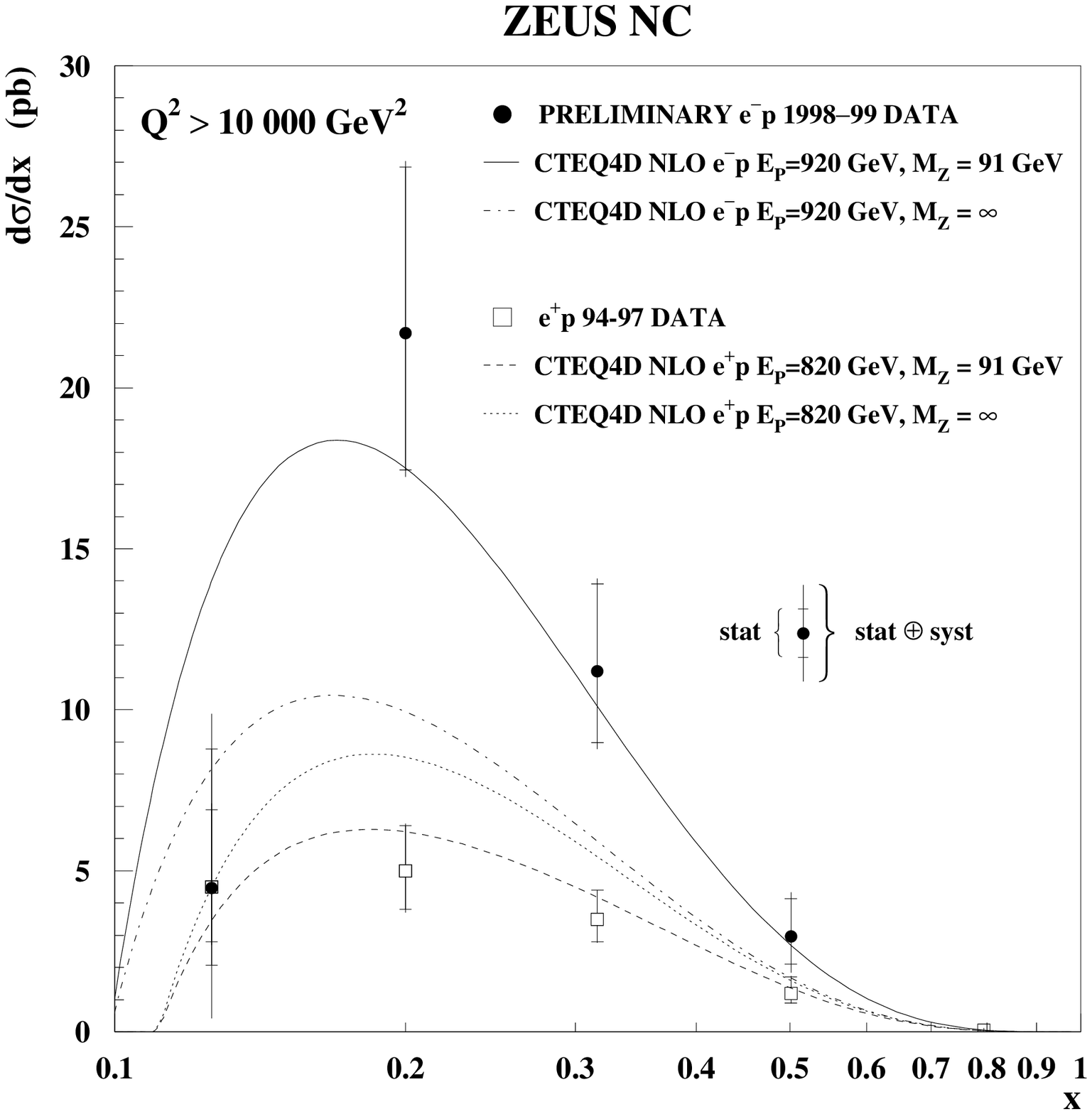,height=6.0cm, width=6.0cm}
\put(-130,130){\makebox(0,0){\footnotesize c)}}
\epsfig{file=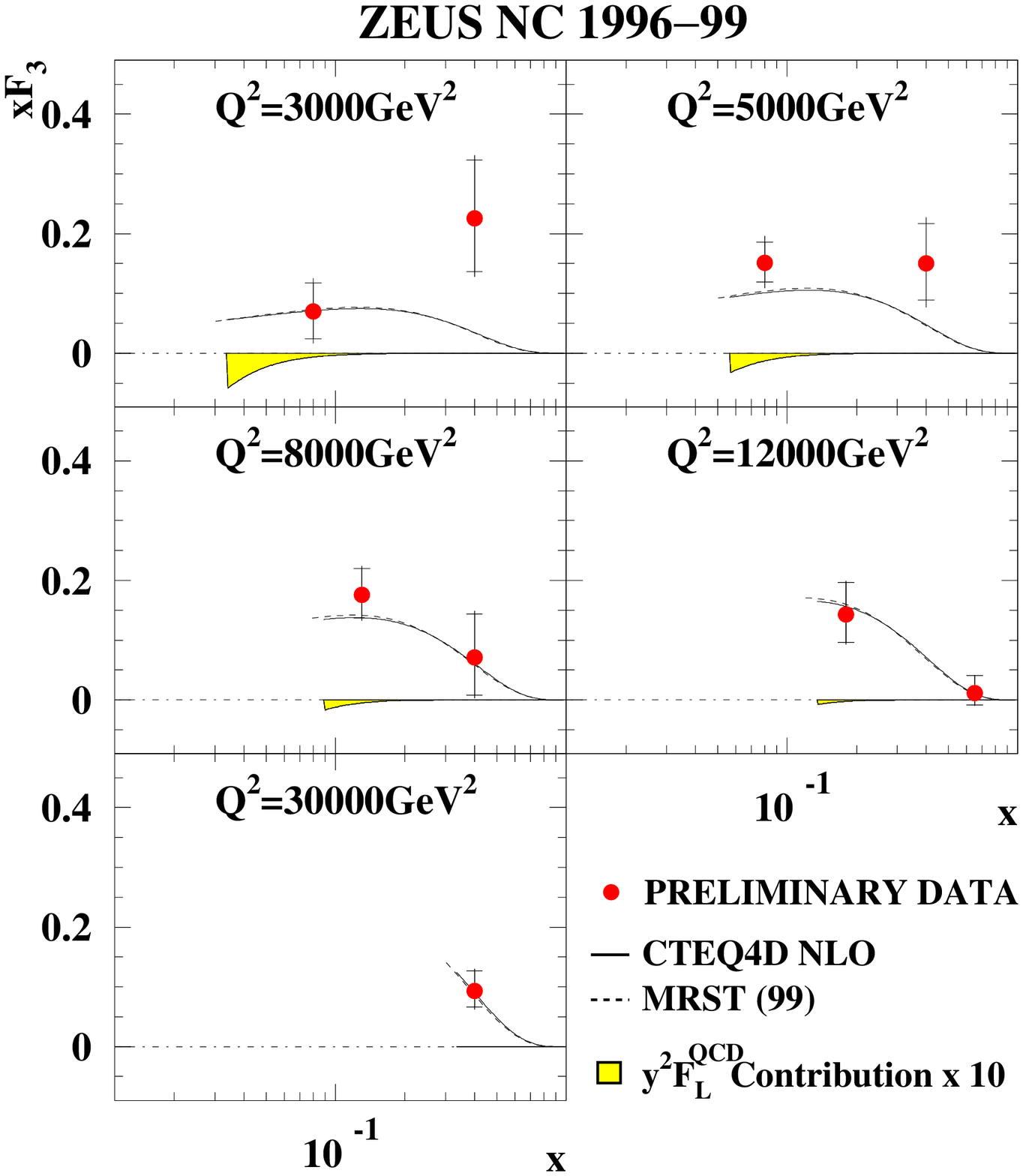,height=6.0cm, width=6.0cm}
\put(-130,140){\makebox(0,0){\footnotesize d)}}
\vspace*{-0.1cm}
    \caption{\footnotesize
      a) The high-$Q^2$ NC $e^-p$ (circles) $e^+p$ (triangles)
reduced cross sections $d^2\sigma_{e^{\pm}p}/dxdQ^2$
from the H1 collaboration along with the Standard Model predictions from
the H1 NLO QCD fit.
      b)The high-$Q^2$ CC reduced cross section $\tilde{\sigma}_{CC}$
from the ZEUS and H1 collaborations along with the Standard Model predictions
using the CTEQ4D parton momentum distributions.
      c) The high-$Q^2$ $e^{\pm}p$ cross section $d\sigma_{e^{\pm}p}/dx$ and
the Standard Model predictions using the CTEQ4D PDFs for $Q^2
> 10\,000 \, \rm{GeV^2}$. Also shown are the theoretical
predictions without the weak interaction were absent ($M_Z =\inf$).
      d) The Structure function $xF_3$, extracted by the ZEUS collaboration.
the data were obtained from the 1998/99 $e^-p$ data set recorded
at $\sqrt{s}=318 \, \rm{GeV}$ and the $e^+p$ data set taken
$\sqrt{s}=300 \, \rm{GeV}$, plotted at fixed $Q^2$ between $3\,000
\, \rm{GeV^2}$ and $30\,000 \, \rm{GeV^2}$ as a function of $x$
together with the Standard Model predictions using the CTEQ4D and
MRST(99) PDFs. 
\label{fig:ddiff}}
\end{center}
\end{figure}

The NC cross-section $d\sigma_{NC}/dx$ for $Q^2 > 10\, 000 \,
\rm{GeV^2}$ is shown in Fig. \ref{fig:ddiff} along with the
expectations from the Standard Model with a $Z^0$ mass of set to the PDG value,
which gives a good description of the data. Also
shown is the expectation from the Standard Model with $Z^0$ mass
set to infinity, thereby effectively removing the weak interaction
so that only photon ($\gamma$) exchange is possible. It can be
seen the data clearly favours the need for the $Z^0$ contribution.

\section{Double Differential NC and CC Cross-Sections}

The double differential reduced NC cross-section is shown in Fig.
\ref{fig:ddiff} for fixed $x$ as a function of $Q^2$ for both
$e^+$ and $e^-$ data. Also shown are the expectations of the
Standard Model which give a good description of the data. For
$Q^2$ values below $~1\, 000 \, \rm{GeV^2}$ the $e^+$ and $e^-$
cross-sections are within errors equal, for higher values of $Q^2$
the cross sections are seen to deviate, with the $e^-p$
cross-section seen to increase above the $e^+p$ cross section.
This difference is consistent with the change in sign of the
parity violating structure function $xF_3$.

The ZEUS collaboration go a step further and combine the $e^+$ and
$e^-$ cross sections, where $30 \, \rm{pb^{-1}}$ of $e^+p$ data
collected at a centre of mass energy of $\sqrt{s}=300 \, \rm{GeV}$
were combined with $16 \, \rm{pb^{-1}}$ of $e^+p$ data collected
at a centre of mass energy of $\sqrt{s}=300 \, \rm{GeV}$. To
reduce statistical fluctuations several bins from the double
differential binning \cite{zeusosaka}. Figure \ref{fig:ddiff}
shows $xF_3$ at fixed values of $Q^2$ as a function of $x$. The
Standard Model expectation evaluated with either the CTEQ4D or
MRST(99) \cite{mrst} parton density functions are seen to give a good
description of the data.


The reduced cross sections as functions of $x$ and $Q^2$ are
displayed in Figs. \ref{fig:ddiff} along with the prediction of the H1 NLO
QCD fit \cite{h1qcd}. The relative increase in the $e^-p$ cross
section over the $e^+p$ cross section is consistent with a larger
$u$ quark density relative to that of the $d$ quark.

The absolute magnitude of the CC cross section, described by equation \ref{eqn:cc} is
determined by the Fermi constant $G_F$ and the PDFs, while the $Q^2$
dependance of the CC cross-section includes the propagator term
$[M_W^2/(M_W^2+Q^2)]$ which produces substantial damping of the cross section.
The ZEUS collabiration have fitted the measured differential cross section,
$d\sigma/dQ^2$, treating $G_F$ and $M_W$ as free parameters, yields

\begin{equation}
G_F=(1.171 \pm 0.034 (\rm(stat.)^{+0.026}_{-0.015}(syst.)^{+0.016}_{-0.015}(PDF)) \times 10^{-5} \, \rm{GeV^{-2}}
\end{equation}

and

\begin{equation}
M_W = 80.8^{+4.9}_{-4.5}(stat.)^{+5.0}_{-4.3}(syst.)^{+1.4}_{-1.3} (PDF) \, \rm{GeV}
\end{equation}

The central values are obtained using the CTEQ4D PDFs. Further fits to the
data were also performed \cite{zeuscc}, the 'propagator-mass' fit to the
measured differential cross-section, $d\sigma/dQ^2$, with $G_F$ fixed to
the value $G_F=1.16639 \times 10^{-5} \, \rm{GeV^{-2}}$ yields the result

\begin{equation}
M_W= 81.4^{+2.7}_{-2.6}(stat.)\pm 2.0 (syst.)^{+3.3}_{-3.0} (PDF) \, \rm{GeV}
\end{equation}

Using the Standard Model constraint $\alpha$, $M_Z$, and all fermion masses,
other than the mass of the top quark, $M_t$, are set to the PDG values
\cite{pdg}. The central result of the fit was obtained with
$M_t = 175 \, \rm{GeV}$ and the mass of the Higgs boson
$M_H = 100 \, \rm{GeV}$. The $\chi^2$ function is evaluated along the line
given by the SM constraint, gives the following result.

\begin{equation}
M_W=80.50^{+0.24}_{-0.25}(stat.)^{+0.13}_{-0.16}(syst.)\pm 0.31(PDF)^{+0.03}_{-0.06} (\Delta M_t,\Delta M_H,\Delta M_Z) \, \rm{GeV}
\end{equation}

\section{HERA Upgrade}

From September 2000, the luminosity of the HERA collider will be increased
by a factor of five. At the same time longitudinal lepton beam polarisation
will be provided for the collider experiments H1 and ZEUS. Over a six year
running period it is anticipated that the total luminosity of
$1\,000 \, \rm{pb^{-1}}$ will be delivered. This large data volume will allow
$F_2^{NC}$ to be extracted with an accuracy of $~3 \%$ over the kinematic
range $2 \times 10^{-5} < x < 0.7$ and $2 \times 10^{-5} < Q^2 < 5
\times 10^4 \, \rm{GeV^2}$. With this accuracy it is anticipated that it
will be possible to determine $\alpha_S$ from the scaling violations of
$F_2^{NC}$ with a precision of $< 0.003$. The gluon distribution will also
be determined from such a fit with a precision of $~3 \%$ for $x=10^{-4}$
and $Q^2 = 20 \, \rm{GeV^2}$.

There will also be a significant benefit to the CC cross-section measurement,
with the potential of a clean determination of the $u$ and $d$ quark densities.

With the introduction of polarised beams it will be possible to obtain a
measurement of the vector and axial-vector couplings of the $u$-quark,
$\nu_u$ and $a_u$ respectively, obtained in a fit in which $\nu_u$ and $a_u$
are allowed to vary. With a luminosity of $250 \, \rm{pb^{-1}}$ per charge,
polarisation combination and taking the vector and axial-vector couplings of
the $u$ and $d$-quarks as free parameters gives a precision of 13 \%, 6 \%,
17 \% for $\nu_u$, $a_u $, $\nu_d$ and $a_d$ respectively. By comparing the
NC couplings of the $c$ and $b$-quarks obtained at LEP, it will be possible
to test the universality of the NC couplings.

\newpage
\section{Summary and Outlook}

The latest Deep-Inelastic Neutral and Charge current cross-sections have
been presented from the H1 and ZEUS experiments. The Standard Model is seen
to give a good description of the data in all cases. The high luminosity
data at high $Q^2$ from both experiments has enabled the first tests of
electroweak effects in both Neutral and Charged current interactions. In NC
interactions the data are seen to be consistent with the effect of $\gamma Z^{\circ}$ interference and has allowed the first extraction of the parity
violating structure function $xF_3$ at high $Q^2$.

With the factor of five increase in yearly luminosity expected
from the HERA upgrade further high precision tests of the Strong
and Weak Interaction will be made. The future CC measurements will
allow precision determinations of the $u$ and $d$ quark densities
and a determination of the vector and axial-vector couplings of
the quarks from the NC interaction, providing important
information for future high energy hadron colliders such as the LHC.

\end{document}